\documentclass[aps,pra,twocolumn,floatfix]{revtex4}
\usepackage{graphicx}
\usepackage{bm}
\usepackage{lscape,graphicx}
\usepackage[dvips]{epsfig}
\usepackage{amsmath}

\begin{document}
\title{Bichromatically driven double well: parametric 
perspective of the control landscape}
\author{Astha Sethi and Srihari Keshavamurthy}
\affiliation{Department of Chemistry, Indian Institute of Technology, 
Kanpur, India 208 016}

\begin{abstract}
We numerically construct and study the 
control landscape of a $(\omega,2\omega)$ bichromatically driven double well 
in the presence of strong fields.
The control landscape is obtained by 
correlating the overlap intensities between the floquet states and
an initial phase space coherent state
with the parametric motion of the quasienergies {\it i.e.,} intensity-level
velocity correlator. 
``Walls" of no control,
robust under variations of the relative phase between the fields,   
are seen in the control landscape and 
associated with multilevel interactions involving chaotic floquet states. 
\end{abstract}

\maketitle

The driven double well system is an important model and an ideal testbed
for several schemes that have been suggested for controlling coherent
superposition of quantum
states\cite{gh98,bsbook,gdjh91,Holth92,fm93,sgsj04}. 
Starting with the
demonstration of enhanced tunneling by Lin and Ballentine\cite{lb90}
and coherent destruction of tunneling\cite{gdjh91} (CDT)
by Grossman {\it et al.} there has been a sustained interest
in the driven double well due to its relevance  
in various areas like quantum computing\cite{qc}, 
coherent control\cite{bsbook}, 
and quantum dots\cite{tkp07}. 
Many of the proposed control scenario 
involve few, carefully chosen, levels which
provide valuable insights into the process.
For example, 
the essence of CDT can be captured
with a two-level floquet state analysis\cite{gp92}.
However, such few-level schemes might be compromised due to several
reasons\cite{cdb07,gb05,lgw94,nr05} and one of them, 
of interest in this work, has to do with the
effect of chaos in the underlying phase space\cite{gb05}.
Typically, control in the presence of strong
fields requires one to go beyond two (or few) level analysis
since floquet states delocalized over the chaotic
region of the classical phase space 
are expected to play an important role\cite{gb05,lgw94,nr05}.
Examples include work 
on bichromatically driven pendulum\cite{lgw94}
and a recent analysis 
of the adiabatic passage technique for coherent population
transfer\cite{nr05}. 

Despite the progress made towards understanding control in the
presence of chaos, a comprehensive understanding of control in terms of  
the sensitivity to initial states, field strengths, and relative
phases is still absent.
The crucial object that one needs for such detailed understanding is
the ``control landscape" since
the structure of the control landscape, amongst other things, 
can shed light on the robustness 
of the control strategies\cite{rhr04}.
To explain what we mean by a control landscape consider 
the Hamiltonian ($\hbar=1$)
\begin{eqnarray}
H(x,p;t)&=&\frac{1}{2m}p^{2}+Bx^{4}-Dx^{2}+xf(t) \label{bichromham} \\
f(t)&=&\lambda_{1} \cos(\omega t) + \lambda_{2} \cos(2\omega t+\phi) \nonumber
\end{eqnarray}
representing a driven double well. 
Our focus on bichromatic control is motivated by extensive studies 
that have shown the utility of such fields in addition to
the possibility of the relative phase $\phi$ providing an additional
control knob\cite{hcu07}.
Owing to the periodicity of Eq.~\ref{bichromham} 
the dynamics of an initial state 
$|z\rangle$ can be analyzed\cite{pg92} in terms of
the floquet states $\{|\chi_{n} \rangle\}$ and the associated
quasienergies $\{E_{n}\}$.
The control landscape for fixed $\omega$ and
a given $|z\rangle$ is the two-parameter space
$(\lambda_{1},\lambda_{2})\equiv \bm{\lambda}$ exhibiting regions of control 
or lack thereof. 
An early example is the $(\lambda_{1},\omega)$ control plane
for CDT which
has been extended to
construct the landscape for a bichromatically 
driven two-level system\cite{dbm96}.
Recently\cite{sgsj04}, 
the difference between the quasienergies
of the floquet states connected to the bare tunneling states
of Eq.~\ref{bichromham} has been used 
to map the control landscape.

\begin{figure} [hbtp]
\includegraphics[height=80mm,width=80mm]{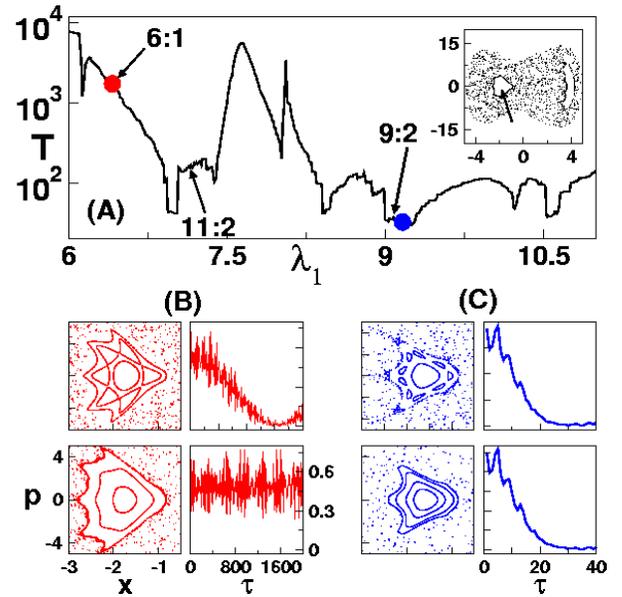}
\caption{(Color online) (A) Decay time $T$ versus
the field strength $\lambda_{1}$ for the coherent state,
$|z\rangle \equiv |x_{0},p_{0}\rangle = |-1.5,0\rangle$,
localized in the left regular island (indicated by an arrow in
the inset showing the stroboscopic phase space at $\lambda_{1}=6.4$) and
$\lambda_{2}=0$. Note the strong fluctuations in $T$ with the plateau
regions around $\lambda_{1}=7.4$, and $9.2$.
(B) and (C) top row shows the local phase space structure and survival
probability for representative points $\lambda_{1}=6.4$ and $9.2$
respectively. The bottom row shows the
effect of the weak control field (relative phase $\phi=0$)
with strength $\lambda_{2}=0.2$ (B) and $\lambda_{2}=0.4$ (C) on the
decay of $|z\rangle$.}
\label{fig1}
\end{figure}

The examples mentioned above clearly
underscore the crucial role of avoided (exact) crossings between
the floquet states. For strong driving fields the
sheer number of such crossings, involving multiple
floquet states, calls for an appropriate measure to
construct the control landscape.    
Here we propose such a measure
based on correlating the
parametric motion of the floquet quasienergy levels (level velocities)
with the overlap intensities 
$p_{zn} \equiv |\langle z|\chi_{n} \rangle|^{2}$.
The overlap intensity-level velocity correlator was introduced
by Tomsovic\cite{Toms96}, followed by detailed studies\cite{kct02}, 
as a sensitive measure 
of localization in phase space.
We show, using Eq.~\ref{bichromham} as example, 
that the correlator is an ideal measure for
constructing and interpreting, perhaps quantitatively, the
control landscape in the presence of strong fields. 

In order to emphasize the role of chaos, consider the  
control problem of Eq.~\ref{bichromham} with
fixed parameter values\cite{lb90,fm93} 
$m=1, D=10, B=0.5$ and, $\omega=6.07$.
The double well supports about eight tunneling doublets 
and $\omega$ nearly corresponds to the energy separation
between the ground and first excited states, neglecting
the tunnel splittings.
Since we are focusing on strong-field control,
we choose $\lambda_{1} \in [6.0,11.0]$ for which the phase space
shows two symmetry-related regular islands embedded
in a chaotic sea.
Earlier, Farrelly and Milligan 
have shown\cite{fm93} that control and suppression
of tunneling can be achieved 
with $\lambda_{2} < \lambda_{1}=10$.
In Fig.~\ref{fig1}(A) the 
decay time $T$, in units of $2\pi/\omega$, of a coherent
state $|z\rangle$ localized in the  
left regular island (cf. Fig.~\ref{fig1}(A) inset)
is shown as a function
of $\lambda_{1}$ for $\lambda_{2}=0$. 
The decay time is infered from
the first vanishing of the  survival probability
\begin{equation}
S(\tau) \equiv |\langle z|\psi(\tau) \rangle|^{2} =
\sum_{m,n}^{} p_{zn}p_{zm} e^{-i(E_{n}-E_{m})\tau}
\label{survprob}
\end{equation}
with the overlap intensities being
$p_{zn} \equiv |\langle z|\chi_{n} \rangle|^{2}$. 
Despite the similar nature of the phase space,
strong fluctuations in $T$ reflect the complicated dynamics
of $|z\rangle$ as determined by the participation of floquet states
and their phase space nature.

A natural question in the context of bichromatic control is as follows.
For a given $T(\lambda_{1},\lambda_{2}=0)$,  
as in Fig.~\ref{fig1}(A), 
what choice of the $2\omega$-field
strength $\lambda_{2}$ and phase $\phi$ will suppress the decay?
Motivated by recent studies\cite{mes06} one suspects that the 
local phase phase structure about $|z\rangle$
is an important factor. 
To this end, in Fig.~\ref{fig1}(B) and (C)
we show the local phase space structure near the left regular island
and the associated $S(\tau)$ for two representative values of $\lambda_{1}$.
In case (B) a prominent $6$:$1$ nonlinear resonance is observed. On
addition of a weak second field $\lambda_{2}=0.2$ the nonlinear resonance
disappears and $S(\tau)$ does not decay to zero. On the other hand
in case (C) a $9$:$2$ resonance is observed for $\lambda_{2}=0$ which
vanishes at $\lambda_{2}=0.4$. However $S(\tau)$ is
hardly effected although in both cases 
the discrete symmetry is broken ($\phi=0$)
upon addition of the $2\omega$-field.  
Similar insensitivity to the local phase space structure occurs in
the vicinity of regions exhibiting plateaus in Fig.~\ref{fig1}(A).
{\em In other words, attempts to control the decay of
$|z\rangle$ in the plateau regions is bound to be difficult}.
Such plateau regions, near regular-chaotic avoided crossings,
have been observed\cite{tu94} in 
the context of chaos-assisted tunneling. Thus, chaotic
floquet states are expected to be key for understanding the features
in Fig.~\ref{fig1}(A). 

\begin{figure} [hbtp]
\includegraphics[height=65mm,width=65mm]{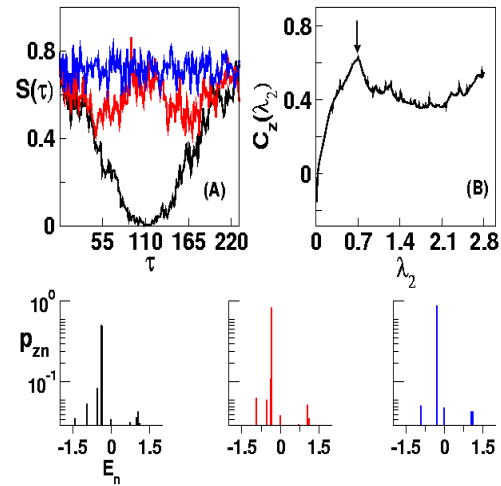}
\caption{(Color online) (A) Survival probability $S(\tau)$
of $|z\rangle = |-1.5,0\rangle$ with
$\lambda_{1}=10$, $\phi=0$ and varying
$\lambda_{2}=0$ (black), $0.7$ (red), and $2.5$ (blue).
(B) The overlap intensity-level velocity correlator (cf. Eq.~\ref{intvelcor})
is shown as a function of $\lambda_{2}$.
$C_{z}(\bm{\lambda})$ exhibits a peak
around $\lambda_{2}=0.7$ indicating substantial localization
and agreeing with the behaviour of $S(\tau)$ in (A).
(Bottom panel) Overlap intensities
for $\lambda_{2}=0,0.7$ and, $2.5$ (left to right)
indicate breaking of the symmetry for $\lambda_{2} \neq 0$.}
\label{fig2}
\end{figure}

Several questions arise at this juncture. What is the precise role
of the chaotic states for control? Are the
plateaus robust for varying $\lambda_{2}$ and $\phi$?
In order to address these questions we motivate the 
intensity-level velocity correlator\cite{Toms96} and show that 
a simple criterion on the correlator highlights
the regimes of control.
The floquet states and quasienergies of  
Eq.~\ref{bichromham} are parametrically dependent on
the field strengths {\it i.e.,} $E_{n}=E_{n}(\bm{\lambda})$ and
$|\chi_{n}\rangle = |\chi_{n}(\bm{\lambda})\rangle$.  
Thus the response of a floquet state to the $2\omega$-field is
measured by the parametric derivative, also known as level velocity, 
$\partial E_{n}/\partial \lambda_{2}$.
At the same time the decay dynamics of a  
coherent state $|z\rangle$ is dominated by
floquet states that have appreciable 
overlap with $|z\rangle$ (cf. Eq.~\ref{survprob}).
Therefore, floquet states $|\chi_{n}\rangle$ with substantial overlap
$p_{zn}$ and large $\partial E_{n}/\partial \lambda_{2}$
are expected to be important in controlling the dynamics of $|z\rangle$. 
The qualitative argument can be made quantitative by introducing the
overlap intensity-level velocity correlator\cite{Toms96} 
\begin{equation}
C_{z}(\lambda_{1},\lambda_{2}) = \frac{1}{\sigma_{z}\sigma_{E}}
\left \langle p_{zn} \frac{\partial E_{n}(\bm{\lambda})}
{\partial \lambda_{2}} \right \rangle_{n}
\label{intvelcor}
\end{equation}
where $\sigma_{z}^2$ and $\sigma_{E}^2$ are the local
variances of $p_{zn}$ and $\partial E_{n}/\partial \lambda_{2}$ respectively.
The average in Eq.~\ref{intvelcor} is
over all the floquet states but the dominant
contributions are expected to come from a finite number of them according
to the qualitative argument above.
In order to keep the
level velocities to be zero centered, the mean is
subtracted and $C_{z}(\bm{\lambda})$ is rescaled to
a unitless quantity with unit variance, making it a
true correlation coefficient\cite{Toms96}. 
In effect,
$C_{z}(\bm{\lambda})$ provides information on groups of
states exhibiting common localization characteristics and is thus
a quantitative measure for deviations from ergodicity.

Utility of the correlator 
can be illustrated with
the example discussed previously\cite{fm93} with
$\lambda_{1}=10$ (fixed), $\lambda_{2}=2.5$ and, $\phi=0$. 
For the chosen parameters the phase space symmetry is broken
and the suppression of tunneling is evident from Fig.~\ref{fig2}(A).
However, Fig.~\ref{fig2}(B) shows that $C_{z}(\bm{\lambda})$ as a function
of $\lambda_{2}$ rises rapidly for small $\lambda_{2}$, 
peaks at $\lambda_{2} \approx 0.7$, and varies slowly for
larger values. Interestingly, similar values
of $C_{z}(\bm{\lambda})$ at
$\lambda_{2}=0.7$ and $\lambda_{2}=2.5$ suggests
comparable extent of localization in phase space. Indeed Fig.~\ref{fig2}(A)
shows that the tunneling is significantly suppressed 
already around $\lambda_{2}=0.7$.
Note that for $\lambda_{2}=2.5$ the symmetry
between the two regular islands is broken markedly resulting in
the considerable suppression of tunneling seen in Fig.~\ref{fig2}(A).
In contrast, for $\lambda_{2}=0.7$ the symmetry breaking is
hardly perceptible in the phase space and yet there is a strong effect
on the quantum dynamics. 
This has been noted\cite{lgw94} before 
and Fig.~\ref{fig2} shows that the correlator is
capable of capturing such effects.

\begin{figure} [hbtp]
\includegraphics[height=90mm,width=90mm]{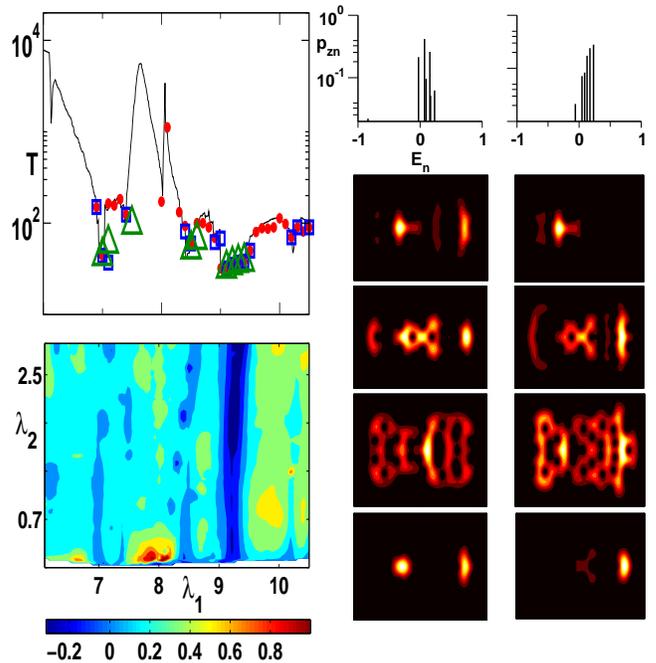}
\caption{(Color online) The left column shows the control landscape (bottom
panel) compared to the decay time plot (top panel, solid line 
as in Fig.~\ref{fig1}(A)).
In the decay time plot only points that are least affected
upon adding the $2\omega$-field
with strengths $\lambda_{2}=0.2$ (red circles), $0.7$ (blue squares),
and $2.1$ (green triangles) are shown.
Note that the regions where there is a lack of control
correspond to the plateaus
with $C_{z}(\bm{\lambda}) \leq 0$.
The middle and right columns show the overlap intensities for
$\lambda_{2}=0$ and $2.1$ respectively with $\lambda_{1}=9.2$ (fixed).
Multiple state interactions involving chaotic floquet states
are involved. The Husimi representations\cite{lb90} of the states are
shown below the corresponding intensity plots with the axis range being
identical to that of the inset in Fig.~\ref{fig1}(A). 
See text for details.}
\label{fig3}
\end{figure}

\begin{figure} [hbtp]
\includegraphics[height=35mm,width=90mm]{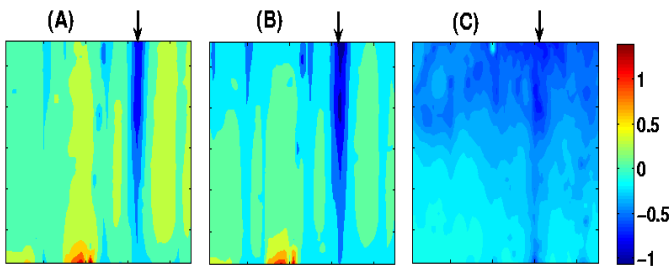}
\caption{(Color online) Control landscape (axes range as in
Fig.~\ref{fig3}) corresponding to
(A) $\phi=\pi/4$, (B) $\pi/3$, and (C) $\pi/2$.
In (C) most of the landscape exhibits $C_{z}(\bm{\lambda}) \leq 0$
indicating lack of control. Regions of control gradually emerge in
(B) and (C) (see also Fig.~\ref{fig3} for $\phi=0$).
Note the robustness of the
``wall'' of no control around $\lambda_{1} \approx 9.2$
(indicated by arrows).}
\label{fig4}
\end{figure}

We now construct the control landscape for 
the entire range of $\lambda_{1}$ shown in Fig.~\ref{fig1}(A).
Since $2\omega$-field is the control,
the intensity-velocity correlator in Eq.~\ref{intvelcor} is evaluated
at a given point $(\lambda_{1},\lambda_{2})$ using the response
of the floquet states with changing field strength $\lambda_{2}$.
The resulting control landscape is shown in Fig.~\ref{fig3} together with the
decay time data of Fig.~\ref{fig1}(A) for relative phase $\phi=0$.
The decay time plot shows regions which
cannot be controlled by the $2\omega$-field which, interestingly, 
correspond precisely to the plateaus in Fig.~\ref{fig1}(A).
A noteworthy feature of the control landscape 
is the existence of ``walls" of no control which are characterized by
$C_{z}(\bm{\lambda}) \leq 0$. In particular, such a wall
at $\lambda_{1} \approx 9.2$ seems to be quite robust and persists for
fairly large values of $\lambda_{2}$. Other such regions around
$\lambda_{1} \approx 7.0$, and $8.2$ seem to break up for higher values
of $\lambda_{2}$. Our analysis reveals that
in the wall regions the dynamics of 
$|z\rangle$ is dominated by multiple floquet states delocalized in
the phase space. 
An example of the multiple floquet contributions, as evident from the
overlaps, and the chaotic nature of
the associated Husimi representations\cite{lb90} is shown in
Fig.~\ref{fig3}. Note that the participation of the chaotic floquet states
at $\lambda_{2}=0$ (no control field) persists even for large $\lambda_{2}$.
Hence, using the symmetry breaking property of the
$2\omega$-field for control purposes is not very effective when
chaotic floquet states are participating in the dynamics. 
On the other hand regions 
with $C_{z}(\bm{\lambda}) > 0$ indicate varying
extent of control of the decay. It is interesting to note that the
ealier analysis\cite{fm93} happens to be in a region
away from such walls of no control!

The effect of varying the relative phase $\phi$ on the control landscape
is shown in Fig.~\ref{fig4}.
The case $\phi=\pi/2$ is
shown in Fig.~\ref{fig4}(C) and one observes that
$C_{z}(\bm{\lambda}) \leq 0$ over most of the landscape which
indicates very little control. This is consistent with the fact
that the combined driving field is symmetric for $\phi=\pi/2$. 
The control landscape for $\phi=\pi/3$, and $\pi/4$ are also shown
in Fig.~\ref{fig4}(B) and (A) respectively. Note that decreasing $\phi$
from the symmetric value of $\pi/2$ gradually leads to regions of control.
However, it is intriguing to see that the wall of no control seen
in Fig.~\ref{fig3} for $\lambda_{1} \approx 9.2$
is present for all the values of $\phi$
shown here. 
For systems with effective $\hbar \ll 1$
one expects several such plateau regions and the effectiveness of the control
would depend sensitively on the initial state and field parameters.

In conclusion, we have constructed the control landscape
for the bichromatically driven double well in the presence of
strong fields using a novel and highly sensitive measure.
A simple criterion on the correlator is able to distinguish
between regions of control and regions of no control.
Lack of control is associated with the involvement of
chaotic floquet states and the
robustness of such regions of no control 
reitrates\cite{cdb07} the need to
excercise care in proposing finite-level control schemes in the
presence of strong fields.
We are currently exploring the usefulness of the correlator, applicable
in the presence of adiabatic fields as well,  
as a tool in formulating local control strategies.

SK gratefully acknowledges useful 
discussions with Prof. Harshawardhan Wanare.
Astha Sethi is funded by a Fellowship 
from the University Grants Commission, India.

\end{document}